\documentclass[aps,prc,reprint,amsmath]{revtex4-1}

\usepackage{amssymb,graphicx}

\newcommand\etal{\textit{~et~al.}}

\begin{document}

\title{Closed expression for the pair vibrational correlation energy
  of a uniform distribution of single-nucleon levels}

\author{K. Neerg\aa rd}
\affiliation{Fjordtoften 17, 4700 N\ae stved, Denmark}

\begin{abstract}
  A closed expression is derived for the pair vibrational correlation
  energy generated in the random phase approximation by the isovector
  pairing force in the case when Kramers and charge degenerate
  single-nucleon levels are uniformly distributed in an interval. The
  expression is used to analyze the spectral density of pair
  vibrational frequencies relative to that of two-quasinucleon
  energies. Applications to the analysis of the symmetry energy of the
  isovector pairing model and to a Strutinskij renormalization of this
  model are discussed.
\end{abstract}

\maketitle

\section{\label{sec:intr}Introduction}

The separable pairing force is a schematic representation of a part of
the interaction of nucleons in the nuclear medium. It was introduced
by Belyayev~\cite{ref:Bel59} in the wake of the adaption to nuclei by
Bohr, Mottelson, and Pines~\cite{ref:Boh58},
Bogolyubov~\cite{ref:Bog58a}, and
Solov'yov~\cite{ref:Sol58a,*ref:Sol58b} of the theory of
superconductivity of Bardeen, Cooper, and
Schrieffer~\cite{ref:Bar57a,*ref:Bar57b}. Its most common application
is in the framework of the nuclear Bardeen-Cooper-Schrieffer (BCS)
theory, where it is supposed to act on otherwise independent nucleons
in a valence space. The Hartree-Bogolyubov approximation is applied to
this Hamiltonian, the pairing Hamiltonian, which amounts in this case
to neglecting, in terms of the general Hartree-Fock-Bogolyubov
scheme~\cite{ref:Bog58b}, the contribution of the pairing force to the
self-consistent single-nucleon potential. The nuclear BCS theory
explains many observations, including the even-odd mass
staggering~\cite{ref:Bel59}, the gap in the spectra of doubly even
nuclei~\cite{ref:Boh58}, moments of
inertia~\cite{ref:Bel59,ref:Mig59}, and enhanced cross sections for
two-nucleon transfer~\cite{ref:Yosh62}.

There was from the outset an interest in exploring the correlations
induced by the pairing force beyond the Hartree-Bogolyubov
approximation. In an early study, Feldman thus diagonalized
numerically the pairing Hamiltonian in a simple case~\cite{ref:Hog61},
Richardson found a reduction of the exact diagonalization of the
Hamiltonian to the solution of a system of non-linear
equations~\cite{ref:Rich63}, and B\`es and Broglia~\cite{ref:Bes66}
used the random phase approximation (RPA)~\cite{ref:Bom53}. This
latter approach was inspired by Bohr's suggestion~\cite{ref:Boh64}
that the pair field might vibrate in a way that is analogous to the
vibrations of the single-nucleon potential accompanying surface
vibrations.

In all this and much later work, separate pairing forces were assumed
to act on neutrons and protons. This interaction is not charge
invariant. The minimal charge invariant extension includes a separable
interaction of isovector pairs of a neutron and a proton. The coupling
constants of the three components must be equal. The RPA was applied
to the resulting, so-called isovector pairing, model in the Sixties by
Ginocchio and Wesener~\cite{ref:Gin68} and recently by
me~\cite{ref:Nee02,ref:Nee03,ref:Nee09}. (My model includes a
schematic interaction of isospins. As this only contributes an energy
proportional to $T(T+1)-\frac34 A$, where $A$ is the mass number and
$T$ the total isospin~\cite{ref:Nee09}, the calculation is equivalent
to one employing the bare isovector pairing force.) Later the
Richardson scheme was extended to the isovector pairing model by
Dukelsky\etal~\cite{ref:Duk06} in a calculation of three levels in
$^{64}$Ge employing the valence space between the magic numbers 20 and
50. Numeric diagonalization of the isovector pairing Hamiltonian in
valence spaces including six or seven Kramers and charge degenerate
single-nucleon levels was done by Bentley and
Frauendorf~\cite{ref:Ben13}. (In both these works the Hamiltonian
includes an interaction of isospins of the same form as that of
Refs.~\cite{ref:Nee02,ref:Nee03,ref:Nee09}.)

Various single-nucleon spectra are employed in my calculations in
Refs.~\cite{ref:Nee02,ref:Nee03,ref:Nee09}. In Ref.~\cite{ref:Nee09}
the levels are generated by a Woods-Saxon potential, while in
Refs.~\cite{ref:Nee02,ref:Nee03} they are equidistant, forming a
so-called picket-fence spectrum. Exploring an equidistant spectrum
aims at displaying average effects of the RPA correlations. To
eliminate in this context the dependence of the results on the valence
space dimension I consider in Ref.~\cite{ref:Nee09}, besides the
Woods-Saxon spectra, a practically infinite picket-fence spectrum.
This approach has the disadvantage that when the BCS gap parameter is
fixed, the RPA energy goes to minus infinity as the valence space
dimension goes to infinity; only the symmetry energy $E_\text{sym} =
E(A,T) - E(A,0)$, where $E(A,T)$ is the total energy, stays finite. I
here approach the aim of displaying average effects of the RPA
correlations in a different manner: The finite picket-fence spectrum
is replaced by a continuous spectrum in an interval. Strutinskij
previously derived in this way a closed expression for the average BCS
energy~\cite{ref:Str67}. Similarly I here obtain a closed expression
for the average RPA energy.

One application of these closed expressions is in calculations such as
those of Bentley, Frauendorf, and me in Ref.~\cite{ref:Ben14}, to
provide smooth counterterms for a Strutinskij renormalization of the
isovector pairing model. Preliminary versions of the present
expressions, communicated in Ref.~\cite{ref:Ben14} without their
derivations, were used in this way in Ref.~\cite{ref:Ben14}. Another
application is demonstrated in Sec.~\ref{sec:spec}. As shown there,
the expression for the RPA energy provides information on the spectral
density of pair vibrational frequencies. It also allows analysis of
the contribution to the symmetry energy of the non-collective pair
vibrational modes in a general way. As discussed in
Refs.~\cite{ref:Nee02,ref:Nee03,ref:Nee09} this contribution
influences the shape of the so-called Wigner cusp in the plot of
masses along a chain of isobaric nuclei. This is the topic of
Sec.~\ref{sec:sym}.

It may be noted finally that several
studies~\cite{ref:Gin68,ref:Ban70,ref:Duk99,ref:Hun07,ref:Ben14} show
the Hartree-Bogolyubov plus RPA to reproduce very accurately the exact
ground state energies of the pairing and isovector pairing
Hamiltonians. For the latter, this approximation is shown, in
particular, in Refs.~\cite{ref:Gin68,ref:Ben14} to be asymptotically
exact in the limit of the coupling constant going to infinity. The
largest deviations occur for values of the coupling constant near
criticality for the onset of a BCS solution with a non-vanishing pair
gap parameter in the case that the critical value is not zero, which
occurs when the Fermi level lies in an interval between consecutive
single-nucleon levels. As the critical value is of the order of the
length of this interval, it vanishes for a continuous spectrum. For
such a spectrum the BCS gap parameter is thus nonzero down to
vanishing of the coupling constant.

As the isovector pairing Hamiltonian is the special case of the
Hamiltonian studied in Ref.~\cite{ref:Nee09} without the so-called
symmetry force, I refer throughout in the following to that article
for details of the formalism. Omitting the symmetry force amounts to
setting there $\kappa = 0$.

The plan of the present article is the following. In
Sec.~\ref{sec:BCS} I review the derivation of Strutinskij's
expression~\cite{ref:Str67} for the BCS energy of a uniform
distribution of single-nucleon levels in an interval. This serves to
set some notation and give some background for the main discussion in
Sec.~\ref{sec:RPA} of the RPA energy generated by this spectrum. In
Sec.~\ref{sec:spec} I use the closed expression obtained in
Sec.~\ref{sec:RPA} to analyze the distribution of RPA frequencies
relative to that of the two-quasinucleon energies. I then turn to the
application of the isovector pairing model to the estimate of nuclear
masses. After a discussion in Sec.~\ref{sec:param} of numeric
parameters I analyze in Sec·~\ref{sec:sym} the contributions to the
symmetry energy of each of the independent-nucleon, BCS, and RPA
energies. Finally, before summarizing the article in
Sec.~\ref{sec:sum}, I discuss the application of the closed expression
derived in Sec.~\ref{sec:RPA} to a Strutinskij renormalization of the
RPA energy of the isovector pairing model.

\section{\label{sec:BCS}BCS}

For a general single-nucleon spectrum the BCS energy $E_\text{BCS}$ is
the difference of the Hartree-Bololyubov energy $E_\text{HB}$ given by
Eq.~(19) of Ref.~\cite{ref:Nee09} (with $\kappa = 0$) and the sum of
occupied single-nucleon levels. For a doubly even nucleus it consists
of a neutron part $E_{\text{BCS},n}$ and a proton part
$E_{\text{BCS},p}$, each given by
\begin{equation}~\label{eq:EBCS}
  E_{\text{BCS},\tau} = 2  \sum_k v_{k\tau}^2 \epsilon_k
    - \frac { \Delta_\tau^2 } G - 2 \sum_{k \le N_\tau / 2} \epsilon_k .
\end{equation}
Here $\epsilon_k$ are the Kramers and charge degenerate single-nucleon
levels and
\begin{equation}\label{eq:uv}
  \left. \begin{matrix} u_{k\tau} \\ v_{k\tau} \end{matrix} \right\}
    = \sqrt{ \frac12 \left( 1 \pm 
      \frac { \epsilon_k - \lambda_\tau } { E_{k\tau} } \right) }
\end{equation}
with
\begin{equation}\label{eq:E}
 E_{k\tau} = \sqrt { (\epsilon_k - \lambda_\tau )^2 + \Delta_\tau^2 } .
\end{equation}
The chemical potential $\lambda_\tau$ and gap parameter $\Delta_\tau$
are determined uniquely by the equations
\begin{equation}\label{eq:lamDel}
  2 \sum_k v_{k\tau}^2 = N_\tau , \quad
  \sum_k \frac 1 { E_{k\tau} } = \frac 2 G ,
\end{equation}
if these equations have a solution. Here $G$ is the pair coupling
constant and $N_n = N$ and $N_p = Z$ are the numbers of neutrons and
protons. These are understood as the numbers of such nucleons
occupying states within the valence space, so they may differ from the
true numbers if a limited valence space is employed. It may happen
that Eqs.~\eqref{eq:lamDel} have no solution; then
$E_{\text{BCS},\tau} = 0$.

I now assume that the single-nucleon levels $\epsilon_k$ are
equidistant with a spacing $1/g$, and that for each $\tau$ a number
$\Omega_{\tau\tau}$ of these levels are selected for the action of the
isovector pairing force on pairs of nucleons of the kind $\tau$. The
selection is assumed symmetric about a level $\lambda_{\tau\tau}$
which turns out equal to $\lambda_\tau$. The interaction of pairs of a
neutron and a proton is passive in the BCS approximation. In the RPA
this is no longer the case. I therefore, in order to prepare the
discussion in Sec.~\ref{sec:RPA}, consider also a selection of a
number $\Omega_{np}$ of single-nucleon levels for the action of the
neutron-proton pairing force. This selection is supposed symmetric
about a level $\lambda_{np}$ which turns out equal to %
$(\lambda_n + \lambda_p)/2$. Assuming each of the three components of
the isovector pairing force to act on selections of single-nucleon
levels that are symmetric about the respective chemical potentials is
the single simplification made in this article, which allows me to
obtain closed expressions for both the BCS and the RPA energy in the
continuous limit. For $N \ne Z$ it implies a deviation from the
isobaric invariance of the original Hamiltonian. The simplification
might be justified by the expectation that details of the
single-nucleon spectrum far from $\lambda_{\tau\tau'}$ should have
little influence on these correlation energies. The three cases
$\tau\tau'=nn$, $pp$ and $np$ are discussed in a unified manner in the
rest of this section, and I drop the index $\tau\tau'$ when it can be
done unambiguously. In the following thus %
$\Omega = \Omega_{\tau\tau'}$ and $\lambda = \lambda_{\tau\tau'}$.
Other quantities introduced in the course of the discussion should
also be understood as specific for the case of $\tau\tau'$. For
convenience in the subsequent analysis, $\Omega$ is supposed to be
always even. The modifications required if $\Omega$ is odd will be
evident.

The continuous approximation results from replacing the sums in
Eqs.~\eqref{eq:EBCS} and~\eqref{eq:lamDel} by integrals. With
\begin{equation}\label{eq:eps><}
  \epsilon_{\gtrless} = \lambda
  \pm \frac \Omega { 2 g } ,
\end{equation}
the second Eq.~\eqref{eq:lamDel} then becomes
\begin{equation}\label{eq:sm gap eq}
  \int\limits_{\epsilon_<}^{\epsilon_>}
    \frac { g d \epsilon }
      { \sqrt { ( \epsilon - \lambda_\tau )^2
        + \Delta_\tau^2 } }
  = g ( a_{\tau>} -a_{\tau<} ) = \frac 2 G ,
\end{equation}
where
\begin{equation}\label{eq:a><}
  a_{\tau\gtrless} = \sinh^{-1} \frac
    { \epsilon_\gtrless - \lambda_\tau } { \Delta_\tau } .
\end{equation}
The first Eq.~\eqref{eq:lamDel} takes the form
\begin{equation}
  \int\limits_{\epsilon_<}^{\epsilon_>} \,
       \Biggl( 1 - \frac { \epsilon - \lambda_\tau }
           { \sqrt { ( \epsilon - \lambda_\tau )^2
             + \Delta_\tau^2 } }
         \Biggr) d \epsilon
  = 2 ( \lambda_\tau^0 - \epsilon_< )
\end{equation}
with
\begin{equation}\label{eq:lamun}
  \lambda_\tau^0 = \lambda
    + \frac { N_\tau - N_{\tau'} } { 4 g } ,
\end{equation}
which can be written
\begin{equation}\label{eq:lambar}
  \lambda_\tau - \lambda_\tau^0
  = \frac { \Delta_\tau } 2 ( e^{-a_{\tau>}} -e^{a_{\tau<}} ) .
\end{equation}
It is easily checked that if Eqs.~\eqref{eq:a><} and~\eqref{eq:lambar}
are satisfied by $\lambda_\tau$ and $\Delta_\tau$, they are also
satisfied by $\lambda_{\tau'} = 2 \lambda - \lambda_\tau$ and
$\Delta_{\tau'} = \Delta_\tau$. Thus %
$\lambda_\tau + \lambda_{\tau'} = 2 \lambda$ and $\Delta_\tau =%
\Delta_{\tau'} \mathrel{\mathop:}= \Delta_{\tau\tau'}%
\mathrel{\mathop:}= \Delta$, whence, in turn, %
$a_{\tau\gtrless} = -a_{\tau'\lessgtr}$. The first of these relations
can be written in more detail as %
$\lambda_{\tau\tau} = \lambda_\tau$ and %
$\lambda_{np} = (\lambda_n + \lambda_p)/2$ as anticipated. If %
$\Omega \gg 2 g \Delta$ then Eq.~\eqref{eq:lambar} gives
\begin{equation}\label{eq:lam0appr}
  \lambda_\tau - \lambda_\tau^0 \approx
    \tfrac12 \left( \frac{ 2 g \Delta } { \Omega } \right)^2
    ( \lambda_\tau^0 - \lambda ) ,
\end{equation}
so that $\lambda_\tau = \lambda_\tau^0$ is then a good approximation
for $\tau\tau' = np$. For $\tau = \tau'$ the equation $\lambda_\tau =
\lambda_\tau^0$ holds exactly by $\lambda_\tau = \lambda$ and
Eq.~\eqref{eq:lamun}.

It is convenient to express other quantities by the parameter
\begin{equation}\label{eq:a}
  a = \tfrac12 ( a_{\tau>} - a_{\tau<} )
    = \tfrac12 ( a_{\tau'>} - a_{\tau'<} ) = \frac 1 { g G } .
\end{equation}
The last expression, which follows from Eq.~\eqref{eq:sm gap eq},
shows $a$ to be a dimensionless reciprocal measure of the coupling
constant $G$. Other convenient relations follow from
Eqs.~\eqref{eq:eps><} and \eqref{eq:a><}:
\begin{multline}\label{eq:hyprel}
  \frac \Omega { g } = \Delta
    ( \sinh a_{\tau>} - \sinh a_{\tau<} ) \\
  = 2 \Delta
    \sinh a \cosh \frac { a_{\tau>} + a_{\tau<}} 2 , \\
  \hskip-4em \delta \lambda =
    \lambda_\tau - \lambda_{\tau'}
  = - \Delta
    ( \sinh a_{\tau>} + \sinh a_{\tau<} ) \\
  \hskip-4em = - 2 \Delta
    \cosh a \sinh \frac { a_{\tau>} + a_{\tau<}} 2 .
\end{multline}
The difference $\delta \lambda$ of chemical potentials is a second
parameter whereby I shall express other quantities.
Equations~\eqref{eq:hyprel} give in particular
\begin{equation}\label{eq:smDel}
  \Delta = \frac \Omega { 2 g \sinh a } \sqrt { 1
     - \left( \frac { g \delta \lambda \tanh a } \Omega \right)^2 } ,
\end{equation}
which for $\tau = \tau'$ becomes
\begin{equation}\label{eq:smDellike}
  \Delta = \frac \Omega { 2 g \sinh a } .
\end{equation}

I can now derive Strutinskij's expression for $E_{\text{BCS},\tau}$.
In this case $\tau = \tau'$, so $\lambda = \lambda_\tau$ and %
$\Delta = \Delta_\tau$. By the first Eq.~\eqref{eq:lamDel} one can
replace $\epsilon_k$ by $\epsilon_k - \lambda_\tau$ simultaneously in
the first and last terms on the right of Eq.~\eqref{eq:EBCS}. With the
sums replaced by integrals and $G$ eliminated by %
Eq.~\eqref{eq:sm gap eq} the expression Eq.~\eqref{eq:EBCS} then
becomes
\begin{multline}\label{eq:smEBCS}
  E_{\text{BCS},\tau} = 
    \int\limits_{\epsilon_<}^{\epsilon_>} \,
       \Biggl[ \Biggl( 1 - \frac { \epsilon - \lambda }
         { \sqrt { ( \epsilon - \lambda )^2 + \Delta^2 } } \Biggr)
         ( \epsilon - \lambda ) \\
    - \frac { \Delta^2 }
      { 2 \sqrt { ( \epsilon - \lambda )^2 + \Delta^2 } } \Biggr]
        g d \epsilon \\
    \hskip-7em - 2 \int\limits_{\epsilon_<}^{\lambda}
      ( \epsilon - \lambda ) g d \epsilon \\
    = - \tfrac12  (1 - e^{-2a} ) g \Delta^2
    = - \tfrac12 \Omega \Delta e^{-a} .
\end{multline}
This is equivalent to Eq.~(11) of Ref.~\cite{ref:Str67} except that a
factor $1/2$ must be missing there by mistake. Belyayev derives in the
continuous limit a related expression for the total Hartree-Bogolyubov
energy including the sum of single-nucleon levels occupied for %
$\Delta = 0$~\cite{ref:Bel59}. The expression~\eqref{eq:smEBCS} is
used in Ref.~\cite{ref:Ben14}.

\section{\label{sec:RPA}RPA}

The additional energy arising from the RPA extension of the
Hartree-Bogolyubov approximation is composed of the terms in Eqs.~(35)
and~(38) of Ref.~\cite{ref:Nee09}. The term $c$ given by Eq.~(35),
which stems from reordering of nucleon fields, vanishes when the
valence space is half filled, so only Eq.~(38) needs to be considered.
The resulting energy $E_\text{RPA}$ splits into a neutron part
$E_{\text{RPA},nn}$, a proton part $E_{\text{RPA},pp}$, and a
neutron-proton part $E_{\text{RPA},np}$, each given by
\begin{equation}\label{eq:ERPA}
  E_{\text{RPA},\tau\tau'} = - \frac i {4 \pi}
    \int\limits_{-\infty}^\infty f( \omega ) d \omega
\end{equation}
with
\begin{equation}
  f ( \omega ) = - \sum_{n=1}^\infty \frac 1 n \, \text{tr} \,
    ( \mathsf V \mathsf G_0 ( \omega ) )^n .
\end{equation}
As a reminder I mostly omit an index $\tau\tau'$, so all of
$f(\omega)$, $\lambda$, $\Delta$, etc. are specific to the case of
$\tau\tau'$. In Eq.~\eqref{eq:ERPA} the matrices $\mathsf V$ and
$\mathsf G_0( \omega )$ have dimensions $2\Omega \times 2\Omega$ and
are composed of $2 \times 2$ blocks (cf.\ in Ref.~\cite{ref:Nee09}
Eqs.~(40) and (43) and the equations before Eq.~(28))
\begin{gather}\begin{split}\label{eq:Vkk'}
  \mathsf V_{kk'} = - G \biggl[ &
    \begin{pmatrix}
      -v_{k\tau}v_{k\tau'} \\ u_{k\tau}u_{k\tau'}
    \end{pmatrix}
    \begin{pmatrix}
      u_{k'\tau}u_{k'\tau'} & -v_{k'\tau}v_{k'\tau'}
    \end{pmatrix}
  \\ & +
    \begin{pmatrix}
      u_{k\tau}u_{k\tau'} \\ -v_{k\tau}v_{k\tau'}
    \end{pmatrix}
    \begin{pmatrix}
      -v_{k'\tau}v_{k'\tau'} & u_{k'\tau}u_{k'\tau'}
    \end{pmatrix}
  \biggr] ,
\end{split} \\ \begin{split}\label{eq:G0kk'}
  & \mathsf G_{0,kk'} ( \omega ) = \\
  & \delta_{kk'} \begin{pmatrix}
    0 & \dfrac 1 { \omega - E_{k\tau} - E_{k\tau'} + i\eta } \\
    \dfrac 1 { - \omega - E_{k\tau} - E_{k\tau'} + i\eta } & 0
  \end{pmatrix} ,
\end{split}\end{gather}
where $\eta$ is a positive infinitesimal. Because, as shown in
Ref.~\cite{ref:Nee09}, $f( \omega )$ is proportional to $\omega^{-2}$
for large $\omega$, one can move the integration path in
Eq.~\eqref{eq:ERPA} to the imaginary axis. So far I only assume that
$\omega$ is not real so that the infinitesimal terms in the
denominators in Eq.~\eqref{eq:G0kk'} can be dropped. From
Eqs.~\eqref{eq:Vkk'} and~\eqref{eq:G0kk'} one gets
\begin{equation}\label{eq:VG0toM}
  \text{tr} \,
      ( \mathsf V \mathsf G_0 ( \omega ) )^n
  = (-G)^n \, \text{tr} \, \mathsf M^n
\end{equation}in terms of the $2\times2$ matrix
\begin{equation}\label{eq:M}
  \mathsf M = \sum_k \, \mathsf X_k \begin{pmatrix}
    \dfrac 1 { \omega - E_{k\tau} - E_{k\tau'} } & 0 \\
     0 & \dfrac 1 { - \omega - E_{k\tau} - E_{k\tau'} }
  \end{pmatrix} \mathsf X_k
\end{equation}
with
\begin{equation}
  \mathsf X_k = \begin{pmatrix}
    u_{k\tau}u_{k\tau'} & -v_{k\tau}v_{k\tau'} \\
    -v_{k\tau}v_{k\tau'} & u_{k\tau}u_{k\tau'}
  \end{pmatrix} . 
\end{equation}
Hence
\begin{multline}\label{eq:f2x2}
  f ( \omega ) = - \sum_{n=1}^\infty \frac { (-G)^n } n \,
    \text{tr} \, \mathsf M^n \\
  = \text{tr} \, \log \, ( \mathsf 1 + G \mathsf M )
  = \log\,\det \, ( \mathsf 1 + G \mathsf M ) ,
\end{multline}
where $\mathsf 1$ is the $2 \times 2$ unit matrix.

By the symmetry of the single-nucleon spectrum about $\lambda$, and
because $\Omega$ is even, the indices $k$ form pairs $(k,k')$ such
that $\epsilon_k + \epsilon_{k'} = 2 \lambda$. As then
\begin{gather}
  E_{k'\tau} + E_{k'\tau'} = E_{k\tau} + E_{k\tau'} , \\
  \mathsf X_{k'} = \mathsf X_k
    \begin{pmatrix} 0 & 1 \\ 1 & 0 \end{pmatrix}
  = \begin{pmatrix} 0 & 1 \\ 1 & 0 \end{pmatrix} \mathsf X_k ,
\end{gather}
the matrix between the two occurrences of $\mathsf X_k$ in
Eq.~\eqref{eq:M} can be replaced by the number
\begin{multline}
  \frac12 \left( \dfrac 1 { \omega - E_{k\tau} - E_{k\tau'} }
  + \dfrac 1 { - \omega - E_{k\tau} - E_{k\tau'} } \right) \\
  \qquad = - \dfrac { E_{k\tau} + E_{k\tau'} }
    { ( E_{k\tau} + E_{k\tau'} )^2 - \omega^2 }
\end{multline}
so that the equation becomes
\begin{equation}
  \mathsf M
  = - \sum_k \dfrac { E_{k\tau} + E_{k\tau'} }
    { ( E_{k\tau} + E_{k\tau'} )^2 - \omega^2 } \,
    \mathsf X_k^2 .
\end{equation}
Because the matrix $\mathsf X_k$ is equivalent to
\begin{equation}
  \begin{pmatrix}
    u_{k\tau}u_{k\tau'} - v_{k\tau}v_{k\tau'} & 0 \\
    0 & u_{k\tau}u_{k\tau'} + v_{k\tau}v_{k\tau'}
  \end{pmatrix}
\end{equation}
by the $k$-independent orthogonal transformation
\begin{equation}
  \frac 1 {\sqrt 2} \begin{pmatrix} 1 & 1 \\ 1 & -1 \end{pmatrix} ,
\end{equation}
the matrix $\mathsf X_k^2$ is equivalent to
\begin{widetext}\begin{equation}\label{eq:(uu+-vv)2}
  \begin{pmatrix}
    ( u_{k\tau}u_{k\tau'} - v_{k\tau}v_{k\tau'} )^2 & 0 \\
    0 & ( u_{k\tau}u_{k\tau'} + v_{k\tau}v_{k\tau'} )^2
  \end{pmatrix}
  = \\ \frac 1 { 4 E_{k\tau} E_{k\tau'} }
  \begin{pmatrix}
    ( E_{k\tau} + E_{k\tau'} )^2
      - \delta \lambda^2 - 4 \Delta^2
    & 0 \\  0 &
    ( E_{k\tau} + E_{k\tau'} )^2 - \delta \lambda^2
  \end{pmatrix} .
\end{equation}\end{widetext}
The reduction in Eq.~\eqref{eq:(uu+-vv)2} follows from Eqs.~(48),
(58), and (59) of Ref.~\cite{ref:Nee09}, where $\tau$ and $\tau'$ may
be substituted for $n$ and $p$ in the last two equations. (The reader
is reminded that $\lambda = \lambda_{\tau\tau'} = %
(\lambda_\tau + \lambda_{\tau'})/2$ and $\delta\lambda = \linebreak %
\lambda_\tau - \lambda_{\tau'}$.) Putting everything together one gets
\begin{widetext}\begin{equation}\label{eq:fscalar}
  f ( \omega ) =
    \log \Biggl[ \Biggl( 1 - \frac G 4 \sum_k \dfrac
      { ( E_{k\tau} + E_{k\tau'} ) [ ( E_{k\tau} + E_{k\tau'} )^2
        - \delta \lambda^2 - 4 \Delta^2 ] }
      { E_{k\tau} E_{k\tau'} 
        [ ( E_{k\tau} + E_{k\tau'} )^2 - \omega^2 ] } \Biggr) \\
      \Biggl( 1 - \frac G 4 \sum_k \dfrac
      { ( E_{k\tau} + E_{k\tau'} ) [ ( E_{k\tau} + E_{k\tau'} )^2
        - \delta \lambda^2 ] }
      { E_{k\tau} E_{k\tau'} 
        [ ( E_{k\tau} + E_{k\tau'} )^2 - \omega^2 )^2 ] } \Biggr)
    \Biggr] .
\end{equation}\end{widetext}

Inserting into Eq.~\eqref{eq:fscalar} the expression
\begin{equation}
  1 = \frac G 4 \sum_k \left( \frac 1 { E_{k\tau} }
      + \frac 1 { E_{k'\tau} } \right)
    = \frac G 4 \sum_k \frac { E_{k\tau} + E_{k\tau'} }
      { E_{k\tau} E_{k\tau'} }
\end{equation}
derived from the second Eq.~\eqref{eq:lamDel}, one gets

\begin{multline}\label{eq:fscred}
  f ( \omega ) = \log \Biggl[
      ( \delta \lambda^2 + 4 \Delta^2 - \omega^2 )
      ( \delta \lambda^2 - \omega^2 ) \\
     \Biggl( \frac G 4 \sum_k \dfrac
      { E_{k\tau} + E_{k\tau'} } { E_{k\tau} E_{k\tau'} 
        [ ( E_{k\tau} + E_{k\tau'} )^2 - \omega^2 ] } \Biggr)^2
    \, \Biggr] .
\end{multline}
Now
\begin{equation}\label{eq:get squares}
  \frac 1 { ( E_{k\tau} + E_{k\tau'} )^2 - \omega^2 }
  = \frac { ( E_{k\tau} - E_{k\tau'} )^2 - \omega^2 }
      { ( E_{k\tau}^2 + E_{k\tau'}^2 -\omega^2 )^2
        - 4 E_{k\tau}^2 E_{k\tau'}^2 }
\end{equation}
and
\begin{multline}\label{eq:elim E2}
  \frac { ( E_{k\tau} + E_{k\tau'} )
      [ ( E_{k\tau} - E_{k\tau'} )^2 - \omega^2 ] }
    { E_{k\tau} E_{k\tau'} } \\
  = \frac { E_{k\tau'}^2 - E_{k\tau}^2 - \omega^2 } { E_{k\tau} }
    + \frac { E_{k\tau}^2 - E_{k\tau'}^2 - \omega^2 } { E_{k\tau'} } ,
\end{multline}
where, by the the symmetry in $\tau$ and $\tau'$, the two terms
contribute equally to the sum in Eq.~\eqref{eq:fscred}.
Equation~\eqref{eq:E} gives
\begin{multline}\label{eq:use E}
  \hskip6em E_{k\tau'}^2 - E_{k\tau}^2
     = 2 ( \epsilon_k - \lambda ) \delta \lambda , \\
  \hskip-9em ( E_{k\tau}^2 + E_{k\tau'}^2 - \omega^2 )^2
      - 4 E_{k\tau}^2 E_{k\tau'}^2 \\
    = 4 ( \delta \lambda^2 - \omega^2 )
            ( \epsilon_k - \lambda )^2
        - ( \delta \lambda^2
            + 4 \Delta^2 - \omega^2 ) \omega^2 \\
    = 4 ( \delta \lambda^2 - \omega^2 )
      [ ( \epsilon_k - \lambda )^2 - q^2 ]
\end{multline}
with
\begin{equation}\label{eq:q,r}
  q = \frac { r \omega } 2 , \quad r = \sqrt { \frac
    { \delta \lambda^2 + 4 \Delta^2 - \omega^2 }
    { \delta \lambda^2 - \omega^2 } } .
\end{equation}
 Moreover
\begin{equation}\label{eq:pole term}
  \frac { 2 ( \epsilon_k - \lambda )
      \delta \lambda - \omega^2 }
    { ( \epsilon_k - \lambda )^2 - q^2 }
  = \frac 1 r \left(
    \frac { r \delta \lambda - \omega }
      { \epsilon_k - \lambda - q }
    + \frac { r \delta \lambda + \omega }
      { \epsilon_k - \lambda + q } \right).
\end{equation}
The branch of the square root in Eq.~\eqref{eq:q,r} may be chosen such
that $r$ is positive when $\omega$ is imaginary.

Consider the first term in the parentheses in %
Eq.~\eqref{eq:pole term}. When this and the factor $1/r$ are inserted
through Eqs.~\eqref{eq:get squares}--\eqref{eq:use E}, the denominator
in the general term in the sum in Eq.~\eqref{eq:fscred} receives a
factor
\begin{equation}
  ( \delta \lambda^2 - \omega^2 ) r
  = \sqrt { ( \delta \lambda^2 + 4 \Delta^2 - \omega^2 )
    ( \delta \lambda^2 - \omega^2 ) } ,
\end{equation}
which is canceled by the factors in front of the squared expression in
parentheses. The sum of the remaining factors becomes
\begin{equation}\label{eq:halfsum}
  \tfrac12 ( r \delta \lambda - \omega ) \sum_k \frac 1
    { ( \epsilon_k - \lambda - q ) E_{k\tau} } .
\end{equation}
To arrive at the continuous approximation I replace the sum in this
expression by the integral
\begin{multline}\label{eq:int}
  \int\limits_{\epsilon_<}^{\epsilon_>} \frac
    { g d \epsilon } { ( \epsilon - \lambda - q )
      \sqrt { (\epsilon - \lambda_\tau )^2 + \Delta^2 } } \\
    = \frac { g } { \Delta \cosh \phi }
      \log \dfrac { \sinh \dfrac { a_{\tau>} - \phi} 2
        \cosh \dfrac { a_{\tau<} + \phi} 2 }
        { \cosh \dfrac { a_{\tau>} + \phi} 2
          \sinh \dfrac { a_{\tau<} - \phi} 2 } ,
\end{multline}
where $\phi$ is any root in
\begin{equation}\label{eq:sinhphi}
  2 \Delta \sinh \phi = 2 q - \delta \lambda
  = r \omega - \delta \lambda
\end{equation}
and the branch of the logarithm is defined by $\log 1 = 0$. The root
$\phi$ can be chosen by the second Eq.~\eqref{eq:q,r} such that
\begin{equation}\label{eq:coshphi}
  2 \Delta \cosh \phi = r\delta \lambda - \omega ,
\end{equation}
which brings the expression~\eqref{eq:halfsum} with the sum replaced
by the integral~\eqref{eq:int} on the form
\begin{equation}\label{eq:redint} 
   g \log \frac
     { \sinh \dfrac { a_{\tau>} - \phi} 2
       \cosh \dfrac { a_{\tau<} + \phi} 2 }
     { \cosh \dfrac { a_{\tau>} + \phi} 2
       \sinh \dfrac { a_{\tau<} - \phi} 2 } .
\end{equation}
Including the contribution from the second term in parentheses in
Eq.~\eqref{eq:pole term} amounts to multiplying the argument of the
logarithm by the expression which results from changing the sign of
$\omega$.

By the second Eq.~\eqref{eq:hyprel} one has
\begin{multline}
  4 \Delta \sinh \dfrac { a_{\tau>} - \phi} 2
    \cosh \frac { a_{\tau<} + \phi } 2 \\
  = 2 \Delta \left( \sinh ( a - \phi )
    + \sinh \frac { a_{\tau>} + a_{\tau<}} 2 \right) \\
  = 2 \Delta ( \sinh a \cosh \phi - \cosh a \sinh \phi )
    - \frac { \delta \lambda } { \cosh a } \\
  = \delta \lambda \left(
       r \sinh a + \cosh a - \frac 1 { \cosh a } \right)
     - \omega ( r \cosh a + \sinh a ) \\
  = ( \delta \lambda \tanh a - \omega ) ( r \cosh a + \sinh a ) .
\end{multline}
This gives the numerator of the fraction in Eq.~\eqref{eq:redint}
expanded by $4 \Delta$. The denominator results from interchanging
$a_{\tau>}$ and $a_{\tau<}$, which amounts to changing the sign of
$a$, and the factors from the second term in parentheses in
Eq.~\eqref{eq:pole term} result from changing the sign of $\omega$.
Totally, the factors $\pm \delta \lambda \tanh a \mp \omega$ cancel
out so that the sum of the expression~\eqref{eq:redint} and its
counterpart for the opposite sign of $\omega$ becomes
\begin{equation}\label{eq:red2int}
   2 g \log \frac { r \cosh a + \sinh a }
    { r \cosh a - \sinh a }
   = 4 g \tanh^{-1} \left( \frac 1 r \tanh a \right)
\end{equation}
with the branch of the inverse hyperbolic tangent given by %
$\tanh^{-1} 0 = 0$.

By substituting the expression~\eqref{eq:red2int} for the sum in
Eq.~\eqref{eq:fscred}, remembering that the factors in front of the
squared expression in parentheses were eliminated, and using
Eq.~\eqref{eq:a}, one gets
\begin{equation}\label{eq:fint}
  f ( \omega ) = 2 \log \left[ \frac 1 a
    \tanh^{-1} \left( \frac 1 r \tanh a \right) \right] .
\end{equation}
As $r$ is by the second Eq.~\eqref{eq:q,r} a function of $\omega^2$,
it is sufficient to do the integral in Eq.~\eqref{eq:ERPA} along the
positive imaginary axis, so
\begin{equation}
  E_{\text{RPA},\tau\tau'} = - \frac i {2 \pi}
    \int\limits_0^{i \infty} f( \omega ) d \omega .
\end{equation}

One can bring this relation on a dimensionless form by setting
\begin{equation}\label{eq:l,y}
  \delta \lambda = 2 \Delta l , \quad
  \omega = 2 i \Delta y .
\end{equation}
This gives
\begin{equation}\label{eq:smERPA}
    E_{\text{RPA},\tau\tau'} = \Delta I(a,l)
\end{equation}
with
\begin{equation}\label{eq:I(a,l)}
  I(a,l) = \frac 2 \pi \int\limits_0^\infty \log \left[ \frac 1 a 
    \tanh^{-1} \left( \frac 1 r \tanh a \right) \right] dy ,
\end{equation}
where, by the second Eq.~\eqref{eq:q,r},
\begin{equation}\label{eq:1/r}
  \frac 1 r = \sqrt { \frac { l^2 + y^2} { 1 + l^2 + y^2 } } .
\end{equation}
As the integrand in Eq.~\eqref{eq:I(a,l)} is evidently negative,
$I(a,l)$ is negative. Some special cases of the general
result~\eqref{eq:smERPA} are discussed in
Secs.~\ref{sec:l=0}--\ref{sec:l><0}.

\subsection{\label{sec:l=0}$l = 0$}

\begin{figure}\centering\includegraphics{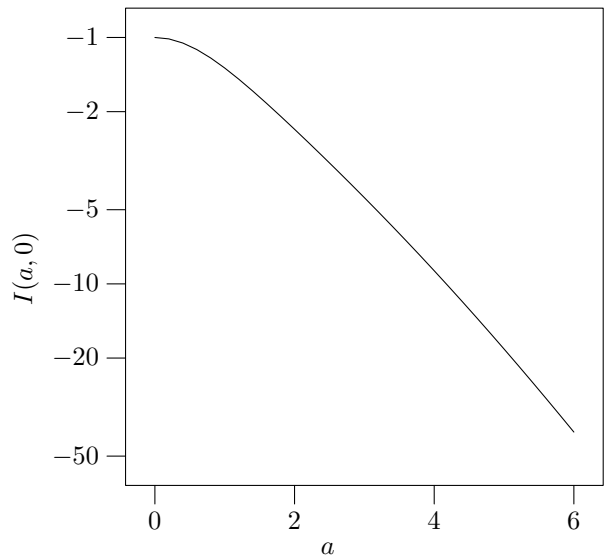}
  \caption{\label{fig:Ia0}The function $I(a,0)$. This function gives
    the two-neutron or two-proton or for $N = Z$ the
    neutron-proton pair vibrational correlation energy in units of
    the gap parameter $\Delta_\tau$. The argument $a$ is the
    reciprocal coupling constant $G$ in units of the single-nucleon
    level spacing $1/g$; see Eq.~\eqref{eq:a}.}
\end{figure} 

\begin{figure}\centering\includegraphics{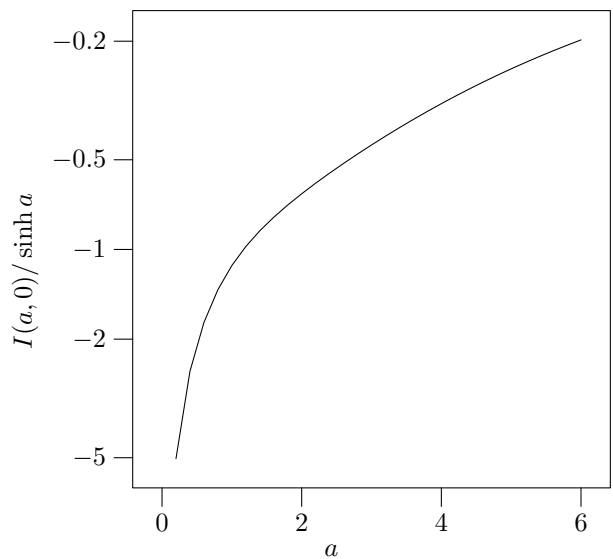}
  \caption{\label{fig:divsinh}The function $I(a,0)/\sinh a$. This
    function is proportional to the two-neutron or two-proton or for
    $N = Z$ the neutron-proton pair vibrational correlation energy for
    a constant single-nucleon level  spacing $1/g$ and valence space
    dimension $4 \Omega$. As to the argument $a$ see the caption to
    Fig.~\ref{fig:Ia0}.}
\end{figure}

For $\tau = \tau'$ and $\tau\tau' =np$ for $N = Z$ one has %
$l = 0$. The substitution
\begin{equation}
  y = \sinh x
\end{equation}
then gives
\begin{equation}
  \frac 1 r = \tanh y ,
\end{equation}
so that Eq.~\eqref{eq:I(a,l)} becomes
\begin{equation}\label{eq:I(a,0)}
  I(a,0) = \frac 2 \pi
    \int\limits_0^\infty \log \left( \frac 1 {2a}
    \log \frac { \cosh ( x + a ) } { \cosh ( x - a ) } \right)
    \cosh x \, dx .
\end{equation}
When inserted in Eq.~\eqref{eq:smERPA}, this gives Eq.~(12) of
Ref.~\cite{ref:Ben14}.

Figure~\ref{fig:Ia0} displays the function $I(a,0)$. It is seen to
decrease rapidly. However, as seen from Fig.~\ref{fig:divsinh},
$I(a,0)/\sinh a$, which by Eqs.~\eqref{eq:smERPA} and \eqref{eq:smDel}
gives the dependence of $E_{\text{RPA},\tau\tau'}$ on $a$ for constant
$g$ and $\Omega$, increases and goes to zero for $a \to \infty$ as
required by Eqs.~\eqref{eq:ERPA}--\eqref{eq:Vkk'} because this limit
corresponds by Eq.~\eqref{eq:a} to $G \to 0$. Figure~\ref{fig:Ia0}
illustrates
\begin{multline}\label{eq:I(0,0)}
  I(0,0) = \frac2\pi \int\limits_0^\infty \log \left[ \frac d {da}
    \tanh^{-1} \left( \frac 1 {r_{l=0}} \tanh a \right)
    \right]_{a = 0} dy \\
  = \frac2\pi \int\limits_0^\infty
      \left( \log \frac 1 {r_{l=0}} \right) dy \\
  = \frac2\pi \int\limits_0^\infty
      \log \left( \frac y { \sqrt {1+y^2} } \right) dy = -1 .
\end{multline}
This limit is not much physically relevant, though, as $a \to 0$
corresponds to $G \to \infty$.

As $\tanh a \to 1$ for $a \to \infty$, the argument of the logarithm
in Eq.~\eqref{eq:I(a,l)} goes to zero in this limit, so %
$I(a,0) \to -\infty$ as illustrated in Fig.~\ref{fig:Ia0}. In
particular, if $\Delta$ is fixed and $a$ determined by
Eq.~\eqref{eq:smDellike} then $\Omega \to \infty$ implies %
$a \to \infty$ and therefore $E_{\text{RPA},\tau\tau'} \to -\infty$ by
Eq.~\eqref{eq:smERPA}. This shows that the exact ground state energy
of the pairing Hamiltonian, which is well approximated by the
Hartree-Bogolyubov plus RPA (see Sec.~\ref{sec:comp}), cannot be
renormalized to a given $\Delta$ in a way approximately independent of
$\Omega$. This is only possible in the BCS approximation, where the
term $e^{-2a}$ in the penultimate expression in Eq.~\eqref{eq:smEBCS}
vanishes for $a \to \infty$.

\subsection{\label{sec:l><0}$l \ne 0$}

\begin{figure}\centering\includegraphics{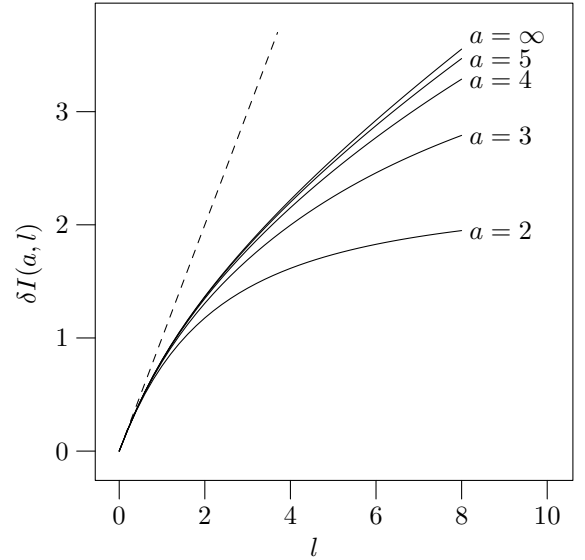}
  \caption{\label{fig:DelI}The function $\delta I(a,l)$ for several
    $a$. This function gives the increment with a neutron excess of
    the neutron-proton pair vibational correlation energy  in units of
    the gap parameter $\Delta_n = \Delta_p$. The argument $l$ is
    the difference $\delta\lambda = \lambda_n - \lambda_p$ of the
    neutron and proton chemical potentials in units of twice the gap
    parameter; see Eq.~\eqref{eq:l,y}. As to the parameter $a$ see the
    caption to Fig.~\ref{fig:Ia0}. The dashed line indicates the
    approximation $\delta I(a,l) = l$.}
\end{figure}

The difference
\begin{multline}\label{eq:DelI}
\delta I(a,l) = I(a,l) - I(a,0) \\
  = \frac 2 \pi \int\limits_0^\infty \log \frac
    { \tanh^{-1} \left(
      \displaystyle \frac 1 r \tanh a \right) }
    { \tanh^{-1} \left(
      \displaystyle \frac 1  {r_{l=0}} \tanh a \right) } dy
\end{multline}
describes the increase, in units of $\Delta$, of $E_{\text{RPA},np}$
with increasing neutron or proton excess. Indeed, by taking
$\lambda_\tau = \lambda_\tau^0$ in the definition of $\delta \lambda$
in Eq.~\eqref{eq:hyprel} one gets from Eq.~\eqref{eq:lamun} and the
first Eq.~\eqref{eq:l,y} that
\begin{equation}\label{eq:l}
  l = \frac { N - Z } { 4 g \Delta }
\end{equation}
for $\tau\tau' = np$. The function~$\delta I(a,l)$, which, evidently
from Eqs.~\eqref{eq:I(a,l)} and \eqref{eq:1/r}, is an even function of
$l$, is plotted for several $a$ in Fig.~\ref{fig:DelI}. As $I(a,l)$ is
negative, $\delta I(a,l)$ necessarily levels off at $-I(a,0)$, which
equals, for example, 2.4 for $a = 2$. As long as $\delta I(a,l)$ is
sufficiently far from this limit, it is seen to be well approximated
by the asymptotic function
\begin{multline}
  \delta I(\infty,l) = \frac 2 \pi \int\limits_0^\infty \log \frac
    { \tanh^{-1} \displaystyle \frac 1 r }
    { \tanh^{-1} \displaystyle \frac 1  {r_{l=0}} } dy \\
  = \frac 2 \pi \int\limits_0^\infty \log \frac
    { \sinh^{-1} \sqrt { l^2 + y^2 } } { \sinh^{-1} y } dy
\end{multline}
and thus nearly independent of $a$.

Substituting $y = l u$ in Eq.~\eqref{eq:DelI} gives
\begin{multline}
  \delta I(a,l) = \\
    \frac { 2 l } \pi \int\limits_0^\infty \log \frac
    { \tanh^{-1} \left( l \tanh a 
      \sqrt { \displaystyle \frac { 1 + u^2 } { 1 + l^2 ( 1 + u^2 ) }
    } \right) }
    { \tanh^{-1} \left( l u \tanh a 
      \sqrt { \displaystyle \frac 1 { 1 + ( l u )^2 ) }
    } \right) } du .
\end{multline}
As
\begin{multline}
\frac
  { \tanh^{-1} \left( l \tanh a 
    \sqrt { \displaystyle \frac { 1 + u^2 } { 1 + l^2 ( 1 + u^2 ) } 
  } \right) }
  { \tanh^{-1} \left( l u \tanh a 
    \sqrt { \displaystyle \frac 1 { 1 + ( l u )^2 ) }
  } \right) }  \\
  \to \frac { \sqrt { 1 + u^2 } } u , \quad l \to 0^+ ,
\end{multline}
the integral becomes in this limit the negative of the one in
Eq.~\eqref{eq:I(0,0)} so that
\begin{equation}\label{eq:linapr}
  \delta I(a,l) \approx l ,
\end{equation}
or,
\begin{equation}\label{eq:part I}
  \left. \frac { \partial I(a,l) } { \partial l} \right|_{l=0^+} = 1 .
\end{equation}
This derivative is illustrated by the dashed line in
Fig.~\ref{fig:DelI}. The result is anticipated because it is
equivalent to
\begin{equation}\label{eq:part E}
  \left. \frac { \partial E_{\text{RPA},np} }
    { \partial \delta \lambda }
    \right|_{ g, \Omega, G,
      \lambda_n + \lambda_p = 2 \lambda,
      \delta \lambda = 0^+} = \tfrac12 .
\end{equation}
For a discrete single-nucleon spectrum the analogous derivative of
$E_{\text{RPA},np}$ with respect to $\delta \lambda$ indeed equals
$1/2$ simply because in Eq.~(39) of Ref.~\cite{ref:Nee09} (also to be
found, for example, in Refs.~\cite{ref:Ban70,ref:Nee03}) the frequency
$|\delta \lambda|$ of a vibrational mode arising from the conservation
of isospin, cf. Ref.~\cite{ref:Gin68} and Sec.~III H of
Ref.~\cite{ref:Nee09}, is the only term in the expression in square
brackets that is not analytic at $\delta \lambda = 0$. This single
frequency continues smoothly into its negative when $\lambda_n$ passes
through $\lambda_p$. I call this mode a \textit{quasi-Goldstone} mode
because it is similar to a Goldstone or Nambu-Goldstone
mode~\cite{ref:Nam60,*ref:Gol61} by arising from a spontaneously
broken symmetry but does not have, in general, zero frequency.

\subsection{\label{sec:comp}Comparison with an exact calculation}

\begin{table*}
  \caption{\label{tbl:comp}Energy in units of the single-nucleon level
    spacing $1/g$ induced by the isovector pairing force in the case
    of a six-level picket-fence spectrum populated by 12 nucleons.
    Shown are the energies calculated by numeric diagonalization of
    the Hamiltonian (exact), in the Hartree-Bogolyubov (HB) plus RPA,
    and in the continuous limit of the latter (continuous). The
    parameter $G$ is the pair coupling constant. I am indebted to Ian
    Bentley for providing the results of numeric diagonalization.}
\begin{ruledtabular}
\begin{tabular}{@{\hspace{2em}}r@{\hspace{2em}}|rrr@{\hspace{2em}}|
  rr@{\hspace{2em}}|rr@{\hspace{2em}}}
& \multicolumn{3}{c|}{$T=0$} & \multicolumn{2}{c|}{$T=2$} &
  \multicolumn{2}{c}{$T=4$} \\
$gG$ & Exact & HB+RPA & Continuous & Exact & HB+RPA & Exact & HB+RPA
\\[2pt]\hline
$0.2$ &  $-2.05$ &  $-2.12$ &  $-2.23$ &  $-1.78$ &  $-1.82$ &
   $-1.51$ &  $-1.54$ \\
$0.4$ &  $-4.65$ &  $-5.11$ &  $-5.05$ &  $-4.02$ &  $-4.27$ &
   $-3.31$ &  $-3.64$ \\
$0.6$ &  $-7.86$ &  $-7.97$ &  $-8.12$ &  $-6.81$ &  $-6.83$ &
   $-5.44$ &  $-5.42$ \\
$0.8$ & $-11.59$ & $-11.59$ & $-11.74$ & $-10.11$ & $-10.07$ &
   $-7.90$ &  $-7.82$ \\
$1.0$ & $-15.74$ & $-15.69$ & $-15.82$ & $-13.81$ & $-13.74$ &
  $-10.60$ & $-10.52$ \\
$1.2$ & $-20.19$ & $-20.12$ & $-20.24$ & $-17.79$ & $-17.71$ &
  $-13.48$ & $-13.40$ \\
$1.4$ & $-24.87$ & $-24.78$ & $-24.89$ & $-21.96$ & $-21.87$ &
  $-16.49$ & $-16.41$ \\
$1.6$ & $-29.69$ & $-29.60$ & $-29.71$ & $-26.26$ & $-26.18$ &
  $-19.58$ & $-19.51$ \\
$1.8$ & $-34.63$ & $-34.54$ & $-34.64$ & $-30.67$ & $-30.58$ &
  $-22.73$ & $-22.67$ \\
$2.0$ & $-39.65$ & $-39.56$ & $-39.66$ & $-35.14$ & $-35.06$ &
  $-25.93$ & $-25.87$
\end{tabular}
\end{ruledtabular}
\end{table*}

A comparison of the results of the Hartree-Bogolyubov plus RPA with
calculations of the exact energy is made in Table~\ref{tbl:comp} in
the case of a six-level picket-fence spectrum populated by 12
nucleons. It is confirmed that the Hartree-Bogolyubov plus RPA is very
good. The largest deviations occur about the minimal $G$ for
nonvanishing gap parameters $\Delta_\tau$, which is given for this
spectrum by $gG \approx 0.35$ almost independently of $T$. This is
explained in Sec.~V of Ref.~\cite{ref:Ben14}. See also
Refs.~\cite{ref:Gin68,ref:Ban70,ref:Duk99,ref:Hun07}.

For $T = 0$ I also show the energies in the continuous limit of the
Hartree-Bogolyubov plus RPA, that is, $E_\text{BCS} + E_\text{RPA}$
with the terms given by Eqs.~\eqref{eq:smEBCS} and \eqref{eq:smERPA}.
This comparison shows that already for six levels in the discrete
picket-fence spectrum the continuous limit is quite representative. I
do not make this comparison for $T > 0$ because the prerequisite of
the derivation in Sec.~\ref{sec:RPA} that for $\tau=\tau'$ the
single-nucleon spectrum be symmetric about the chemical potentials
$\lambda_\tau$ is obviously badly violated for $N - Z = 2T = 4$ and 8
when the valence space includes only six levels.

\section{\label{sec:spec}Spectrum of pair vibrational frequencies}

It follows from the derivation of Eq.~(39) of Ref.~\cite{ref:Nee09}
that the discontinuity of $f( \omega )$ across a branch cut at the
real axis describes the cumulated spectral density of pair vibrational
frequencies $\omega_k$ relative to that of the two-quasinucleon
energies $E_{k\tau} + E_{k\tau'}$. The closed expression
Eq.~\eqref{eq:fint} therefore allows an analysis of this relative
spectral density. Because $f( \omega )$ is an even function of
$\omega$ and real for imaginary $\omega$, complex conjugation of
$\omega$ maps to complex conjugation of $f(\omega)$, so the
discontinuity equals the value of $-i \Im f(\omega) / 2$ immediately
below the real axis. By the derivation in Ref.~\cite{ref:Nee09} this
gives
\begin{equation}
  \Im f ( \omega - i \eta )
    = \pi  \sum_k  [ \theta ( \omega - \omega_k )
      - \theta ( \omega - ( E_{k\tau} + E_{k\tau'} ) ) ]
\end{equation}
for $\omega > 0$.

\begin{figure}\centering\includegraphics{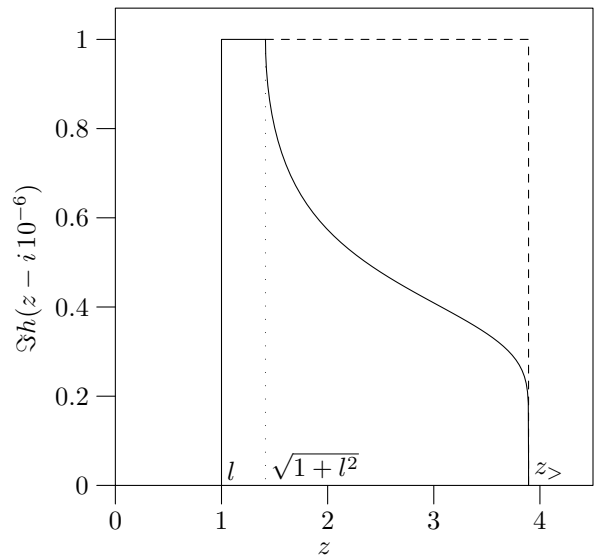}
  \caption{\label{fig:Imf}The function $\Im h( z - i \, 10^{-6} )$ for
    $a,l = 2,1$. This function gives (when $10^{-6}$ is viewed as
    infinitesimal) the cumulated density of the neutron-proton and
    for $l = 0$ the two-neutron and two-proton pair vibrational
    frequencies relative to that of the two-quasinucleon energies. The
    argument $z$ is the vibrational frequency $\omega$  in units of
    twice the gap parameter $\Delta_n = \Delta_p$. The labels on the
    abscissa axis and the dashed line are explained in the text.}
\end{figure}

It is convenient to introduce again a dimensionless measure of
$\omega$. This time I set $\omega = 2 \Delta z$ and define accordingly
\begin{equation}\label{eq:h}
  h(z) = \frac 1 \pi f( 2 \Delta z )
    = \frac 2 \pi \log \left[ \frac 1 a
      \tanh^{-1} \left( \frac 1 r \tanh a \right) \right]
\end{equation}
with
\begin{equation}\label{eq:1/r(z)}
  \frac 1 r = \sqrt { \frac { l^2 - z^2} { 1 + l^2 - z^2 } } .
\end{equation}
Then $\Im h( z - i \eta )$ gives for $z > 0$ directly the cumulated
relative spectral density. An example of this function is plotted in
Fig.~\ref{fig:Imf}.

The shape of the resulting curve is easily understood from the
discussion in Sec.~III~J of Ref.~\cite{ref:Nee09}. The cumulated
relative spectral density $\Im h( z - i \eta )$ jumps from 0 to 1 at
the quasi-Goldstone frequency $z = l$. It may be noticed that this
frequency remains separated from the bulk of the spectrum in the
continuous limit. A second vibrational frequency occurs at %
$z = \sqrt { 1 + l^2 }$ (Eq.~(59) of Ref.~\cite{ref:Nee09}), but
immediately thereafter with increasing $z$ the first twofold
degenerate two-quasinucleon energy appears. In the discrete spectrum,
$\Im h( z - i \eta )$ therefore first rises to 2 and then drops to 0.
As the rest of the vibrational frequencies are also twofold degenerate
and each of them is embedded between sucessive two-quasinucleon
energies, this alternation of 2 and 0 continues until the highest
two-quasinucleon energy, which occurs at $z = z_>$, where
\begin{multline}\label{eq:z>}
  z_>^2 =  \\
  \biggl[ \frac 1  { 2 \Delta } \biggl(
    \sqrt { ( \epsilon_> - \lambda_\tau)^2
      + \Delta^2 }
    + \sqrt { ( \epsilon_> - \lambda_{\tau'})^2
      + \Delta^2 } \,\biggr) \biggr]^2 \\
  = \left( \frac12 ( \cosh a_{\tau>} + \cosh a_{\tau<} ) \right)^2 \\
  = \left( \cosh a \cosh \frac { a_{\tau>} + a_{\tau<}} 2 \right)^2  \\
  = \left( \cosh a \right)^2
    + \left( \cosh a \sinh \frac { a_{\tau>} + a_{\tau<}} 2 \right)^2 \\
  = \left( \cosh a \right)^2 + l^2
\end{multline}
by Eqs.~\eqref{eq:eps><}, \eqref{eq:a><}, \eqref{eq:hyprel}, and
\eqref{eq:l,y}. In the continuous limit $\Im h( z - i \eta )$ becomes
the average of these 2 and 0 weighted by the lengths of the intervals
in which each of them prevails. Figure~\ref{fig:Imf} shows that at the
beginning of the continuous spectrum the vibrational frequencies are
situated midway between consecutive two-quasinucleon energies. With
increasing frequency they then rapidly approach the two-quasinucleon
energy above.

Because the path of integration in Eq.~\eqref{eq:ERPA} can be
transformed as described in Sec.~III~C of Ref.~\cite{ref:Nee09}, the
integral of $\Im h( z - i \eta )$ for $z > 0$ is $-I(a,l)$. The shape
of the plot of $\Im h( z - i \eta )$ as displayed in
Fig.~\ref{fig:Imf} thus provides a deeper understanding of the
behavior of the function $I(a,l)$, including, in particular, its
deviation from linearity in $l$. Thus, if $\Im h( z - i \eta )$ were
replaced by a constant in the interval $l < z < z_<$, as suggested by
the dashed line in Fig.~\ref{fig:Imf}, and $z_>$ were constant, then
the integral would be strictly linear in $l$ due to the displacement
with $l$ of the lower edge of the curve. The displacement of the upper
edge adds a positive term equal to the increase of $\epsilon_>$.
Finally the subtracted area between the solid and dashed curves
shrinks with $z_> - \sqrt{ 1 + l^2 }$, which decreases with increasing
$l$. This adds another positive contribution. Both these contributions
are quadratic in $l$ to the lowest order and give the deviation from
linearity. This argument is seen to provide also an alternative
derivation of Eq.~\eqref{eq:part I}.

The shape of the curve in Fig.~\ref{fig:Imf} is also easily understood
from the expressions~\eqref{eq:h} and \eqref{eq:1/r(z)}. First notice
that the imaginary part of $h(z)$ is $2/\pi$ times the complex
argument of
\begin{equation}\label{eq:phi}
  \phi(z) = \tanh^{-1} \psi(z) , \quad
  \psi(z) = \frac 1 r \tanh a .
\end{equation}
I discuss how this develops with increasing $z > 0$. For $z < l$ the
square root in Eq.~\eqref{eq:1/r(z)} is real and less than one. As
also $\tanh a < 1$, the functions $\psi(z)$ and $\phi(z)$ are real. At
$z = l$ the square root branches off in opposite imaginary values,
the one below the cut being positive. Then $\phi(z)$ is also positive
imaginary, so $\arg \phi(z)$ jumps to $\pi/2$. It stays there until $z
= \sqrt{ 1 + l^2 }$, when the square root becomes positive real again.
At this point, however, $\psi(z)$ is infinite, so $\phi(z)$ has
reached $i \pi / 2$. As $\psi(z)$ then descends from infinity through
positive real values, $\phi(z)$ takes values of an increasing positive
real number plus $i \pi / 2$ by continuity, and so $\arg \phi(z)$
descends. This continues until $r = \tanh a$, when $\Re \phi(z)$ is
infinite so that $\arg \phi(z)$ vanishes. At this point $\phi(z)$ thus
has a pole. Walking around the pole the branches above and below the
cut then join in a real value and $\phi(z)$ stays real. It is easily
shown by Eqs.~\eqref{eq:1/r(z)} and \eqref{eq:z>} that $r = \tanh a$
is equivalent to $z = z_>$.

\section{\label{sec:param}Parameters for numeric estimates}

Four parameters enter the expressions for the BCS and RPA energies
derived in Secs.~\ref{sec:BCS} and \ref{sec:RPA}:
\begin{enumerate}
  \item the number $\Omega$ of Kramers and charge degenerate
    single-nucleon levels supposed to participate in pair
    correlations,
  \item the density $g$ of such levels, understood to pertain to the
    neighborhood of the Fermi level,
  \item the pair coupling constant $G$,
  \item in $E_\text{RPA,$np$}$ for $N \ne Z$, the difference %
    $\delta \lambda$ of neutron and proton chemical potentials.
\end{enumerate}
I discuss the choice of these parameters for the purpose of numeric
estimates.

There is no obvious way to extract $\Omega$ from data. It is
desirable, however, to use a recipe that is the simplest possible,
involves the least possible structural assumptions, and is consistent
with the prerequisite of the derivations in Secs.~\ref{sec:BCS} and
\ref{sec:RPA} that the single-nucleon levels be symmetrically
distributed about the respective chemical potentials. These criteria
are satisfied if in each of the three cases $\tau\tau' = nn$, $pp$,
and $np$ one includes all levels from the bottom of the spectrum to
the Fermi level and equally many levels upwards from there. This
amounts to taking $\Omega = N_\tau$ for $\tau = \tau'$ and %
$\Omega = A/2$ for $\tau\tau' = np$. Incidentally these are also the
approximate numbers of bound levels in the nuclear potential well.

For the single-nucleon level density $g$ I adopt the value extracted
by Kataria, Ramamurthy and Kapoor~\cite{ref:Kat78} from observed
neutron resonances in spherical nuclei,
\begin{equation}\label{eq:g}
  \frac {\pi^2} 6 \times 4 g = 0.176 \text{ MeV}^{-1} ( A - A^{-2/3} ) .
\end{equation}

The pair coupling constant $G$ is generally extracted in some manner
from observed odd-even mass differences. Strutinskij~\cite{ref:Str67}
thus calculates $G$ from Eqs.~\eqref{eq:a} and \eqref{eq:smDel} by
presumably (cf. Ref.~\cite{ref:Bra72}) identifying $\Delta$ with the
odd-even mass difference $\Delta_\text{oe}$ and adopting Bohr's and
Mottelson's fit~\cite{ref:Boh69}
\begin{equation}\label{eq:oefit}
  \Delta_\text{oe} = 12 A^{-1/2} \text{ MeV}
\end{equation}
to the observed values. This procedure makes $G$ somewhat dependent on
$\Omega$. The dependence is seen, however, to be logarithmic. On the
other hand identifying $\Delta$ directly with $\Delta_\text{oe}$ is a
severe simplification. Its rationale is that in the BCS theory,
$\Delta$ is the energy of a quasinucleon excitation at the Fermi
level, cf. Eq.~\eqref{eq:E}. This excitation blocks, however, the Fermi
level from taking part in the pair correlations, thus reducing the
effective density of participating levels. The reduction of the
effective $g$ increases the parameter $a$ by Eq.~\eqref{eq:a} and thus
reduces by Eqs.~\eqref{eq:smDel} and \eqref{eq:smEBCS} and
Fig.~\ref{fig:divsinh} the absolute values of both $E_\text{BCS}$ and
$E_\text{RPA}$. For a reliable determination of $G$ one therefore
needs to do a full calculation of the binding energies of both the
odd-$A$ nuclei and their doubly even neighbours and then fit $G$ to
reproduce the observed differences.

Moreover, twice the expression~\eqref{eq:oefit} is seen from Fig.~3 of
Ref.~\cite{ref:Ben13} or Fig.~6 of Ref.~\cite{ref:Ben14} to
underestimate greatly the observed $T = 0$ doubly odd--doubly even
mass differences above $^{56}$Ni. In Ref.~\cite{ref:Ben14}, Bentley,
Frauendorf and I fit the $T = 0$ doubly odd--doubly even mass
differences from $A = 24$ to $A = 100$ in a full, Strutinskij
renormalized calculation, cf. Sec.~\ref{sec:ren}, based on the
isovector pairing Hamiltonian with the above $\Omega$. We find
\begin{equation}\label{eq:G}
  G = 8.6 A^{-4/5} \text{ MeV}
\end{equation}
to be optimal. I therefore adopt this expression.

Equations~\eqref{eq:a}, \eqref{eq:g} and \eqref{eq:G} give
\begin{equation}\label{eq:aemp}
  a = \frac {4.35} {A^{1/5} - A^{-2/15}} .
\end{equation}
For example $a = 3.5$ for $A = 24$, $a = 2.6$ for $A = 56$ and %
$a = 2.2$ for $A = 100$. On can then infer from Eq.~\eqref{eq:smDel}
that
\begin{equation}
  \frac12 \left( \frac {2 g \Delta} {\Omega} \right)^2
    \le \frac 1 { 2 (\sinh a)^2}\le 0.025 ,
\end{equation}
where the final bound corresponds to $a = 2.2$. It then follows from
Eq.~\eqref{eq:lam0appr} that $\lambda_\tau = \lambda_\tau^0$ is a very
good approximation, improving with smaller values of $a$. In this
approximation
\begin{equation}\label{eq:dellam}
  \delta \lambda = \frac {N -Z} {2 g}
\end{equation}
follows from Eq.~\eqref{eq:lamun}.

\section{\label{sec:sym}Symmetry energy}

I discuss in Refs.~\cite{ref:Nee02,ref:Nee03,ref:Nee09} the symmetry
energy of the isovector pairing model. In this section I examine which
new insights the closed expressions derived in Secs.~\ref{sec:BCS}
and~\ref{sec:RPA} might bring to this discussion. For the modeling of
the symmetry energy it is sufficient to consider the isobaric analog
with the maximal $N$, so $T$ can be identified with $(N - Z)/2$.

In the Hartree-Bogolyubov plus RPA the total energy $E$ includes the
sum $E_\text{indep}$ of single-nucleon levels subtracted in
Eq.~\eqref{eq:EBCS}. Thus
\begin{equation}\label{eq:Etot}
  E = E_\text{indep} + E_\text{BCS} + E_\text{RPA} .
\end{equation}
The term $E_\text{indep}$ is composed of a neutron part
$E_{\text{indep},n}$ and a proton part $E_{\text{indep},p}$, each given by
\begin{equation}
  E_{\text{indep},\tau} = 2 \sum_{k \le N_\tau / 2} \epsilon_k .
\end{equation}
In the continuous limit this becomes
\begin{equation}
  E_{\text{indep},\tau}
    = 2 g \int_{\epsilon_<}^\lambda \epsilon \, d \epsilon
    = g ( \lambda^2 - \epsilon_<^2 ) 
    = \epsilon_< N_\tau + \frac {N_\tau^2} {4 g}
\end{equation}
by Eq.~\eqref{eq:eps><} and $\Omega = N_\tau$. As the filling of the
single-nucleon spectrum from the bottom for both $\tau$ implies
$\epsilon_{n<} = \epsilon_{p<} \mathrel{\mathop:}= \epsilon_<$, adding
the neutron and proton contributions results in
\begin{equation}
   E_\text{indep}
     = \epsilon_< A + \frac {A^2} {8 g} + \frac {T^2} {2 g} .
\end{equation}
The contribution to $E_\text{sym}$ is $T^2/2 g$.

In the approximation~\eqref{eq:linapr} the part $\Delta \delta I(a,l)$
of $E_\text{RPA}$ equals $T/2 g$ by Eqs.~\eqref{eq:smERPA} and
\eqref{eq:l}. It is noteworthy that this term \emph{depends on neither
  $\Omega$ nor $G$}, which are empirically the least well determined
parameters. In combination with the contribution from $E_\text{indep}$
it gives a total term in $E_\text{sym}$ equal to $T(T+1)/2 g$.

It is well known~\cite{ref:Boh69} that $1/2 g$ is much less than the
empirical symmetry energy coefficient. For example, for $A = 56$
Eq.~\eqref{eq:g} gives $1/2g = 0.45$ MeV, while the coefficient of
$T(T+1)$ in the semiempirical mass formula, Eq.~(1) of
Ref.~\cite{ref:Men10} by Mendoza-Temis, Hirsch, and Zucker, is
1.29~MeV. The difference must be attributed to isospin-dependent
interactions~\cite{ref:Boh69}. In
Refs.~\cite{ref:Nee02,ref:Nee03,ref:Nee09} I consider a schematic
two-nucleon interaction, called the symmetry force in
Refs.~\cite{ref:Nee03,ref:Nee09},
\begin{equation}\label{eq:sym}
  V_{12} = \kappa \, \mathbf t_1 \cdot \mathbf t_2 ,
\end{equation}
where $\kappa$ is a coupling constant and $\mathbf t$ is the nucleonic
isospin. The symmetry force contributes an energy
\begin{equation}
  \tfrac12 \kappa [ T(T+1) - \tfrac34 A ] ,
\end{equation}
so that, totally so far,
\begin{equation}\label{eq:Esym}
  E_\text{sym} = \frac12 \left( \frac 1 g + \kappa \right) T(T+1) .
\end{equation}

The most important lesson to be learned from this result is that it
supports taking in semiempirical mass formulas the symmetry energy
proportional to $T(T+1)$, such as done for example by Mendoza-Temis,
Hirsch, and Zucker, rather than proportional to $T^2$ as is more
usual. It should be noticed that this conclusion does not rest on the
present very schematic model where otherwise independent nucleons in a
valence space interact by the isovector pairing and symmetry forces.
It applies to any Hamiltonian which produces a self-consistent
Hartree, Hartree-Fock, Hartree-Bogolyubov, or Hartree-Fock-Bogolyubov
self-consistent state that is not an eigenstate of $N - Z$. If
$E_\text{sc}(T)$ is the self-consistent energy, a generalization of
the discussion in Sec.~III~H of Ref.~\cite{ref:Nee09} implies, in
fact, that the quasi-Goldstone RPA mode restoring isobaric invariance
has frequency $E_\text{sc}'(T)$. The only requirement for this
relation to hold is that the RPA stability matrix, Eq.~(8.73) of Ring
and Schuck~\cite{ref:Rin80}, is the Hessian matrix of $E_\text{sc}$
with respect to variations about self-consistency. Adding to
$E_\text{sc}(T)$ the term in Eq.~(39) of Ref.~\cite{ref:Nee09} from
the quasi-Goldstone frequency gives $E_\text{sc}(T) + \frac12
E_\text{sc}'(T)$. In the neighborhood of $T = 0$ the self-consistent
energy $E_\text{sc}(T)$ rises proportionally to $T^2$, so
$E_\text{sc}(T) + \frac12 E_\text{sc}'(T)$ rises proportionally to
$T(T+1)$.

I now discuss the remaining contributions to $E_\text{sym}$ in the
expression~\eqref{eq:Etot}. Throughout I understand $\Omega = N_\tau$
for $\tau = \tau'$ and $\Omega = A/2$ for %
$\tau\tau' = np$ to be substituted wherever $\Omega$ occurs in
formulas. Equations~\eqref{eq:smDellike} and \eqref{eq:smEBCS} then
give
\begin{equation}
  E_\text{BCS} = - \frac { N^2 + Z^2 } { 4 g (e^{2 a} - 1 ) }
    = - \frac { A^2/4 + T^2 } { 2 g (e^{2 a} - 1 ) } .
\end{equation}
The BCS energy thus generates a small term quadratic in $T$. Its
coefficient $1 / ( 2 g (e^{2 a} - 1 ) )$, is easily estimated to
amount to at most a few permille of the coefficient of $T(T+1)$ in
Eq.~(1) of Ref.~\cite{ref:Men10}. From Eqs.~\eqref{eq:smERPA} and
\eqref{eq:smDellike} one gets
\begin{equation}\label{eq:ERPA0}
    E_{\text{RPA},nn} +E_{\text{RPA},pp}
      = \frac { A I(a,0) } { 2 g \sinh a }
      \mathrel{\mathop:}= 2 E_\text{RPA}^0 ,
\end{equation}
which does not depend on $T$. With $g$ and $a$ given by
Eqs.~\eqref{eq:g} and \eqref{eq:aemp} the energy $E_\text{RPA}^0$
equals $-5.4$ MeV for $A = 24$, $-6.4$ MeV for $A =56$, and $-7.1$ MeV
for $A =100$.

\begin{figure}\centering\includegraphics{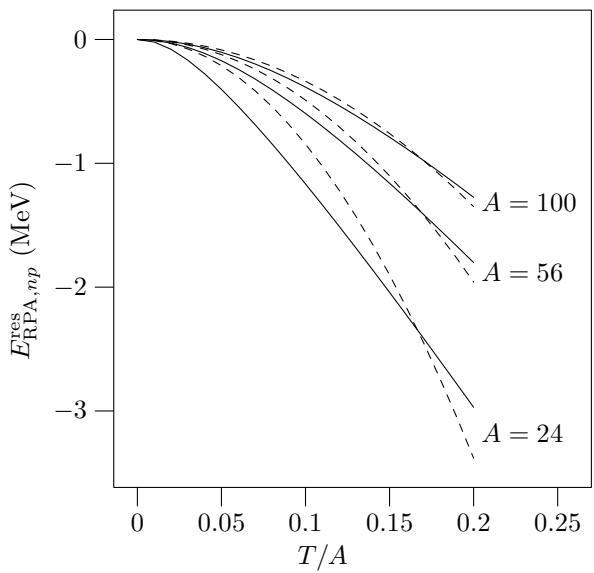}
  \caption{\label{fig:Eres}The residual neutron-proton pair
    vibrational correlation energy $E_{\text{RPA},np}^\text{res}$ as a
    function of $T/A$. The dashed lines show least square fits by
    quadratic functions in the interval of the plot.}
\end{figure}

It remains to discuss the terms in $E_{\text{RPA},np}$ in excess of
$E_\text{RPA}^0$ and the term $T/2 g$ from the quasi-Goldstone RPA
mode. With the square root factor in Eq.~\eqref{eq:smDel} denoted by
$s$ one has
\begin{multline}\label{eq:ERPAres}
  E_{\text{RPA},np}^\text{res}
     = E_{\text{RPA},np} - E_\text{RPA}^0 - \frac T {2 g} \\
     = (s - 1) E_\text{RPA}^0
       + \frac {s A} {4 g \sinh a} (\delta I(a,l) - l )
\end{multline}
with
\begin{equation}
  l = \frac {2 T \sinh a} {s A}
\end{equation}
by Eq.~\eqref{eq:l}. Equation~\eqref{eq:lamun} gives
\begin{equation}\label{eq:s}
  s = \sqrt { 1 - \left( \frac { 2 T \tanh a } A \right)^2 } .
\end{equation}
The two terms in Eq.~\eqref{eq:ERPAres} have opposite signs but the
second one dominates. Figure~\ref{fig:Eres} shows
$E_{\text{RPA},np}^\text{res}$ as a function of $T/A$ for $A = 24$, 56
and 100 and $T < 0.2 A$. It is is seen to take in this range negative
values of the order of a few MeV with the numerically largest values
occurring for the lowest $A$. Also shown are least square fits by
quadratic functions. Their coefficients make up $-5.9$, $-1.2$, and
$-0.4\%$, respectively, of the coefficient of $T(T+1)$ in Eq.~(1) of
Ref.~\cite{ref:Men10}. It follows that in a fit of the symmetry energy
of the isovector pairing plus symmetry force model with a $\kappa$
that reproduces approximately the empirical masses, by a function
proportional to $T(T+X)$, the optimal $X$ is $1/0.941$, $1/0.988$ and
$1/0.996$ respectively. The quadratic approximation is seen to be
poor, however, and poorest for the lowest $A$. Anyway, the deviations
do not exceed some hundred MeV, which is within the accuracy of
semiempirical mass formulas. The bending down of
$E_{\text{RPA},np}(A,T) - E_{\text{RPA},np}(A,0)$ from linearity in
$T$ at high $T$ displayed in Fig.~\ref{fig:Eres} is well known from my
previous studies of discrete single-nucleon spectra. It is quantified
in a simple form by Eq.~\eqref{eq:ERPAres}.

As discussed in Refs.~\cite{ref:Nee02,ref:Nee03,ref:Nee09} the term in
Eq.~\eqref{eq:Esym} linear in $T$ as well as the bending down from
linearity described by the term $E_{\text{RPA},np}^\text{res}$
contribute to the average shape of the Wigner cusp in the plot of
masses along an isobaric chain. In particular the vanishing of
$E_{\text{RPA},np}$ at large $T$ is reminiscent of the behavior of the
phenomenological ``Wigner energy'' of exponential form proposed by
Myers and Swiatecki~\cite{ref:Mye66}. It should be borne in mind,
however, that the contribution from $E_{\text{RPA},np}$ makes up less
than half of the total linear term in Eq.~\eqref{eq:Esym} and that
shell effects contribute very significantly to the Wigner cusp of an
individual isobaric chain~\cite{ref:Nee09,ref:Ben13,ref:Ben14}.

Notice finally that for $T = 0$ the total RPA energy equals $3
E_\text{RPA}^0$. It thus takes values about $(-15)$--$(-20)$ MeV for
$A =24$--100.

\section{\label{sec:ren}Strutinskij renormalization}

The idea of the Strutinskij theory~\cite{ref:Str67} is to view in a
first approximation the nucleus as a liquid drop whose properties may
be derived from semiempirical mass formulas. The deviation of the
actual mass from the liquid drop average is viewed as a ``shell
correction'' which must be calculated microscopically. As only this
small correction needs to be calculated from a microscopic model, the
model need not be very accurate; in the simplest version of the theory
the microscopic energy is just the sum of occupied levels in a
potential well. To calculate the shell correction one must subtract
from the microscopic energy an average that depends smoothly on the
parameters of the model. Replacing this average by the liquid drop
energy is known as a Strutinskij renormalization. The notation of the
present section is such that a symbol without a tilde denotes a
quantity calculated from the microscopic model and the same symbol
with a tilde its smooth counterterm. If $x$ is any quantity, $\delta x
= x - \tilde x$.

The pairing and isovector pairing models are crude models offering
themselves to Strutinskij renormalization. Strutinskij in fact uses
his expression~\eqref{eq:smEBCS} to provide the smooth counterterm for
a renormalization of the of BCS energy. The
expression~\eqref{eq:smERPA} may be applied analogously to renormalize
the RPA energy. It is used in a preliminary form in this way by
Bentley, Frauendorf, and me in Ref.~\cite{ref:Ben14}.

An important role is played in the Strutinskij theory by a smooth
single-nucleon level density $\tilde g$, which is a function of the
single-nucleon energy $\epsilon$. It is obtained by spreading each
microscopic single-nucleon level over an interval of the order of the
distance of the major shells. In terms of this function one can define
smooth chemical potentials $\tilde \lambda_\tau$ by
\begin{equation}
  \int_{-\infty}^{\tilde\lambda_\tau} \tilde g(\epsilon) d \epsilon
    = N_\tau .
\end{equation}
In his calculation~\cite{ref:Str67} of $\tilde E_\text{BCS}$,
Strutinskij uses formulas equivalent to those of Sec.~\ref{sec:BCS}
with $g = \tilde g(\tilde \lambda_\tau)$. The parameter $\Omega$ is
taken as the number of single-nucleon levels included in the
microscopic BCS calculation. The approach to the calculation of
$\tilde E_\text{RPA}$ taken in Ref.~\cite{ref:Ben14} is analogous with
$g = \tilde g(\tilde \lambda_{np})$ in $\tilde E_{\text{RPA},np}$. Here
$\tilde \lambda_{np}$ is defined by
\begin{equation}
  \int_{-\infty}^{\tilde\lambda_{np}} \tilde g(\epsilon) d \epsilon
    = \frac A 2 .
\end{equation}
In these calculations the lowest $A/2$ single-nucleon levels are
included in all parts of the microscopic calculations and $\Omega$
accordingly set to $A/2$ for all $\tau\tau'$ in the calculations of
the counterterms.

At the time of these calculation Eq.~\eqref{eq:I(a,l)} had not been
derived. Only Eq.~\eqref{eq:I(a,0)} was known to us and used for %
$N = Z$. For $N > Z$ we used an approximation which is essentially
equivalent to neglecting $\tilde E_{\text{RPA},np}^\text{res}$, namely
Eq.~(13) of Ref.~\cite{ref:Ben14}, where, unfortunately, a factor 1/2
is missing in the last term by mistake. As seen from
Fig.~\ref{fig:Eres}, $\tilde E_{\text{RPA},np}^\text{res}$ can take
values of minus several MeV. Using Eq.~\eqref{eq:smERPA} diminishes
$\tilde E_\text{RPA}$ by the absolute value of this amount and thus
increases the total energy by the same absolute value. It is noticed
that this change is largest for the smallest $A$ and the largest $T$.
A full calculation by the scheme of Ref.~\cite{ref:Ben14} using
Eq.~\eqref{eq:smERPA} instead of the preliminary $\tilde E_\text{RPA}$
requires a refit to the data of $G$ and the liquid drop parameters.
Such work is in progress.

\begin{figure}\centering\includegraphics{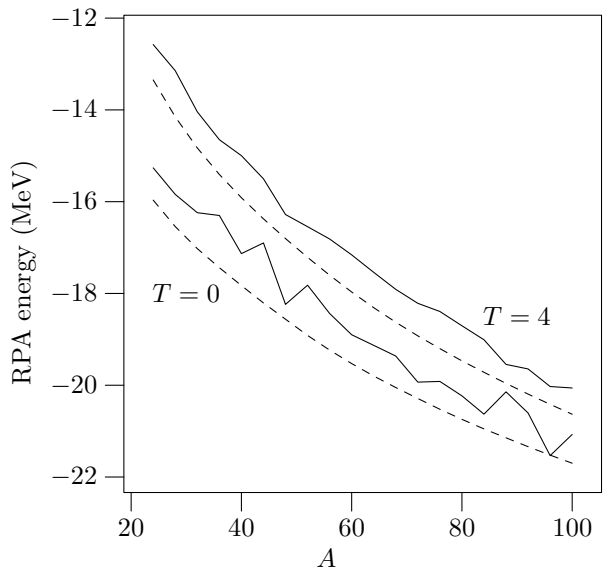}
  \caption{\label{fig:stru}RPA energies of doubly even nuclei as
    functions of $A$. The solid lines shows the RPA energies from
    Ref.~\cite{ref:Ben14} and the dashed curves below each solid line
    are their smooth counterterms given by Eq.~\eqref{eq:smERPA}.}
\end{figure}

In Fig.~\ref{fig:stru} the RPA energies of Ref.~\cite{ref:Ben14} are
compared to their smooth counterterms calculated by
Eq.~\eqref{eq:smERPA}. The RPA correction $\delta E_\text{RPA}$ is
seen to be almost constant about 0.7 MeV. As such a constant term can
largely be absorbed by adjustment of the liquid drop parameters, the
RPA correction thus turns out to have little significance for the
reproduction of the observed doubly even masses. The shell correction
$\delta E_\text{indep} + \delta E_\text{BCS}$ suffices for this
purpose within the general accuracy of the model. It is crucial,
however, for the reproduction of the masses in the vicinity of $N = Z$
and thus, in particular, of the Wigner cusp that the liquid drop
symmetry energy be proportional to $T(T+1)$ rather than $T^2$. The
proportionality to $T(T+1)$ can be motivated microscopically only by
the reasoning in Sec.~\ref{sec:sym}. Moreover an RPA contribution to
the $T = 0$ doubly even--doubly odd mass staggering remains.

The fact that $E_\text{RPA}$ is consistently greater than $\tilde
E_\text{RPA}$ could be understood from the fact that at equilibrium
deformation the effective microscopic single-nucleon density $g$ is
always at the Fermi level lower than $\tilde g(\tilde \lambda)$.
Therefore the effective $a$ is higher by Eq.~\eqref{eq:a} and,
consequently, $E_\text{RPA}$ is less negative by Eq.~\eqref{eq:ERPA0}
and Fig.~\ref{fig:divsinh}.

\null

\section{\label{sec:sum}Summary}

The main result of this article is the closed
expression~\eqref{eq:smERPA} for the pair vibrational correlation
energy generated in the random-phase approximation (RPA) by the
isovector pairing force in the case when Kramers and charge-degenerate
single-nucleon levels are uniformly distributed in an interval. Using
this expression I analyzed the distribution of pair vibrational
frequencies relative to that of two-quasinucleon energies. This
analysis revealed among other results that, like for the previously
studied discrete single-nucleon spectra, quasi-Goldstone pair
vibrational frequencies produced by the breaking of isobaric
invariance by the self-consistent Bogolyubov quasinucleon vacuum are
in the continuous limit isolated from the bulk of the spectrum. The
total distribution of pair vibrational frequencies was found to
account in a simple way for features of the neutron-proton pair
vibrational correlation energy observed both in the previous studies
and presently: a linear increase with the isospin $T$ near $T = 0$ and
a bending down from linearity at higher $T$.

The emergence in the isovector pairing model, possibly amended by a
schematic interaction of isospins, of a symmetry energy proportional
to $T(T+1)$ for low $T$ was reviewed in terms of
Eq.~\eqref{eq:smERPA}, and the universal character of this result as a
consequence of the breaking of isobaric invariance by the
self-consistent state was pointed out. The deviation from
proportionality to $T(T+1)$ at higher $T$ was expressed by a simple
formula and found to be largest for the lowest mass numbers $A$.

Finally the application of Eq.~\eqref{eq:smERPA} to a Strutinskij
renormalization of the RPA energy of the isovector pairing model was
discussed. Significant modifications of the calculated masses were
found to result from using Eq.~\eqref{eq:smERPA} instead of an
approximation to this expression applied in recent calculations by
Bentley, Frauendorf and me. The difference between the microscopic RPA
energy and a smooth counterterm expressed by Eq.~\eqref{eq:smERPA}
turned out to be almost constant about 0.7 MeV. Therefore, upon
Strutinskij renormalization, the RPA contribution is insignificant for
the reproduction of the observed doubly even masses. It is crucial,
however, for such masses near $T = 0$ to be well described that the
proportionality of the symmetry part of the total smooth energy to
$T(T+1)$ rather than $T^2$ for low $T$ be preserved in the replacement
of the smooth energy by a liquid drop energy.

\bibliography{pairvib}

\end{document}